\documentclass[a4paper,12pt]{article}
\usepackage{amsmath,amsfonts,amssymb,cite}
\usepackage{graphicx
}

\begin{document}

\begin{center}
{\LARGE\bf The quark masses and meson spectrum: A holographic
approach}
\end{center}

\begin{center}
{\large S. S. Afonin and I. V. Pusenkov}
\end{center}

\begin{center}
{\small V. A. Fock Department of Theoretical Physics,
Saint-Petersburg
State University, 1 ul. Ulyanovskaya, St. Petersburg, 198504, Russia\\
Email: \texttt{afonin@hep.phys.spbu.ru}
}
\end{center}

\begin{abstract}
The spectrum of radially excited unflavored vector mesons is
relatively well measured, especially in the heavy-quark sector.
This provides a unique opportunity to observe the behavior of the
hadron spectrum at fixed quantum numbers as a function of the
quark mass. The experimental data suggests the approximately Regge
form for the radial spectrum, $M_n^2=An+B$, where $A$ and $B$ are
growing functions of the quark mass. We use the bottom-up
holographic approach to find the functions $A$ and $B$. The
obtained result shows a good agreement with the phenomenology and
consistency with some predictions of the Veneziano-like dual
amplitudes.
\end{abstract}

\section{Introduction}

The bottom-up holographic approach to QCD~\cite{son1,pom} turned
out to be very  interesting and fruitful laboratory for the
theoretical study of the phenomenology of the strong interactions.
Traditionally this approach is applied to the spectroscopy of the
light hadrons and to the description of the related physics (the
low-energy physics, hadron formfactors, finite-temperature effects
etc., see, e.g., Refs.~\cite{rewr,schmidt} for references). Up to
now not much efforts have been invested in the holographic
description of the heavy-quark sector. In particular, we are aware
of only one attempt~\cite{schmidt} to describe analytically the
excited spectrum of heavy hadrons as a function of the quark
masses within the framework of the holographic approach. The
purpose of the present work is to address this problem in the case
of the unflavored vector mesons.

Our choice of the hadron states is driven by the fact that, for
the heavy mesons, only in the unflavored vector case a reach
experimental spectrum of the radial excitations is
available~\cite{pdg}. Since the radial excitations emerge
naturally in the 5D holographic models
--- they are identified with the Kaluza-Klein modes --- the chosen
sector can be tested phenomenologically. In addition, the
holographic description of the vector mesons is relatively
simple~\cite{son1,pom} and looks most naturally as one deals with
the conserved currents.

The $S$-wave unflavored vector mesons are intensively produced in
the $e^+e^-$-annihilation. The mechanism of resonance formation
for such states is expected to be universal at all available
energy scales. We will assume that in the relativistic
picture\footnote{Here the crucial point is that one works with the
boson masses squared. Passing to the linear masses (the
non-relativistic picture) a flavor-dependent "binding energy" will
appear.} the contribution to the mesons masses stemming from the
gluon interactions is flavor-independent within the accuracy of
the holographic approach to be used. The approximate value of this
contribution is given by the spectrum of the $\omega$-mesons in
which the quark masses can be set to zero. In other words, the
spectrum of unflavored vector mesons is assumed to depend on the
quark masses and the other contributions are encoded in the
universal coefficients of the corresponding mass formula.
\begin{table}
\caption{\small The masses of known $\omega$, $\psi$ and
$\Upsilon$ mesons (in MeV)~\cite{pdg}. The experimental error is
not displayed if it is less than 1~MeV. The following least
reliable states are omitted: $\omega(2330)$ (and another candidate
$\omega(2290)$)~\cite{pdg}, all $D$-wave $\psi$-mesons~\cite{pdg}
and also $\Upsilon(11023)$~\cite{pdg} (the last resonance has a
small coupling to the $e^+e^-$-annihilation in comparison with
$\Upsilon(10860)$ --- this suggests a strong $D$-wave admixture in
this resonance).}
\begin{center}
{
\begin{tabular}{|c|ccccc|}
\hline
$n$ & $0$ & $1$ & $2$ & $3$ & $4$\\
\hline
$M_\omega$ & $783$ & $1425 \pm 25$ & $1670 \pm 30$ & $1960 \pm 25$ & $2205 \pm 30$ \\
$M_\psi$ & $3097$ & $3686$ & $4039 \pm 1$ & $4421 \pm 4$ & --- \\
$M_\Upsilon$ & $9460$ & $10023$ & $10355$ & $10579 \pm 1$ & $10865 \pm 8$ \\
\hline
\end{tabular}
}
\end{center}
\end{table}
\begin{table}
\caption{\small The radial Regge trajectories~\eqref{1} (in
GeV$^2$) for the data from Table 1 (see text).}
\begin{center}
{
\begin{tabular}{|c|c|}
\hline
$M_n^2$ & $An+B$\\
\hline
$M_\omega^2$ & $0.95n+0.99$\\
$M_\psi^2$ & $2.98n+10.5$\\
$M_\Upsilon^2$ & $5.75n+95.1$\\
\hline
\end{tabular}}
\end{center}
\end{table}

\begin{figure}[ht]
\begin{minipage}[ht]{0.46\linewidth}
\includegraphics[width=1\linewidth]{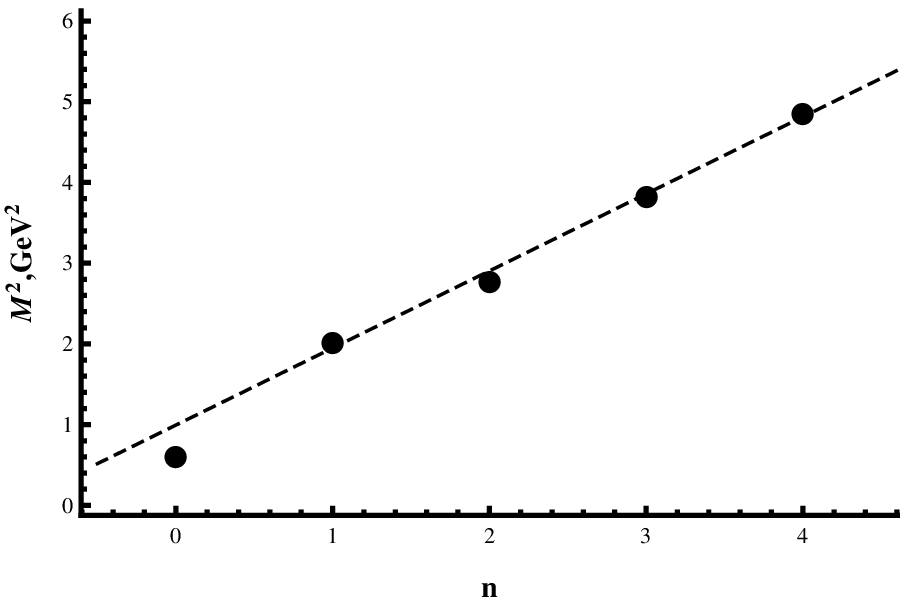} \\
{\scriptsize Fig. (1a). The spectrum of the $\omega$-mesons. The
experimental points (for this and subsequent figures) are taken
from Table 1.}
\end{minipage}
    \hfill
   \begin{minipage}[ht]{0.46\linewidth}
    \includegraphics[width=1\linewidth]{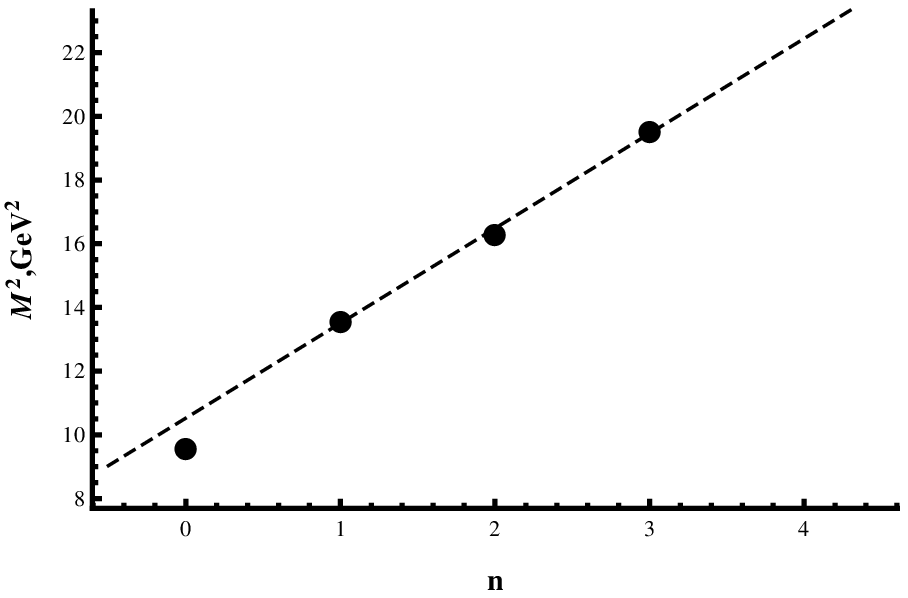} \\
    {\scriptsize Fig. (1b). The spectrum of $\psi$-mesons.}
    \end{minipage}

     \vspace{0.7cm}
\begin{minipage}[ht]{0.46\linewidth}
\includegraphics[width=1\linewidth]{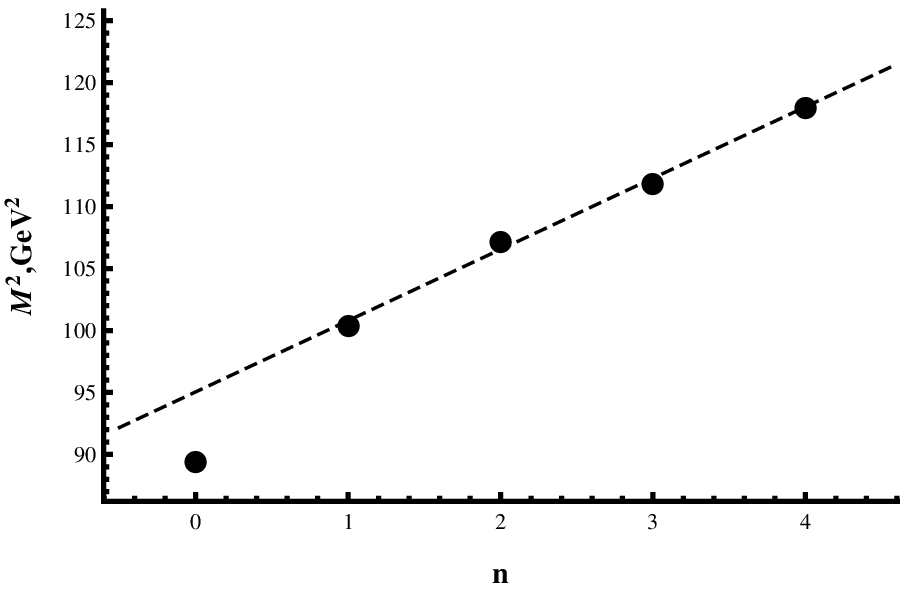} \\
{\scriptsize Fig. (1c). The spectrum of $\Upsilon$-mesons.}
\end{minipage}

\end{figure}

The spectrum of states we are going to describe is given in
Table~1. The Figs.~(1a)--(1c) show that the radial spectrum reveal
a universal Regge-like behavior $M_n^2\sim n$, where $n$ is the
radial quantum number. There exists another universal effect: The
ground states lie systematically below the linear trajectory.
Probably some universal dynamics causes this effect. Likely the
given effect is related with a closer location of the valent
quarks in the ground states than in the excited ones. In the
language of the non-relativistic potential models, this could mean
that the confinement potential is strongly distorted at typical
sizes of the ground states by the Coulomb part, by the spin-spin
or other interactions. We are not aware of any discussions on this
point in the literature. Irrespectively of the physical reason
behind the effect in question, we find reasonable to exclude the
ground states in fitting the corresponding linear trajectories
which we are going to describe holographically. The fits made with
the linear ansatz
\begin{equation}
\label{1}
M_n^2=An+B,
\end{equation}
are displayed in Table~2. It is clearly seen that both the slope
$A$ and the intercept $B$ grow in response to increasing the quark
mass. But the rate of this growth is quite different --- the
intercept grows much faster.

Anticipating some possible questions, an important remark should
be made. Usually the linear Regge-like formula~\eqref{1} is
applied to the interpolation of the spectra of the light mesons
since such a behavior is expected from the semiclassical QCD
string considerations with massless quarks. It is rather
surprising that this linearity holds also for the heavy vector
mesons, with the accuracy of the interpolation~\eqref{1} being the
same or even better. In addition, the last two states on the
trajectory of Fig.~(1a) need confirmation, while all states on
Figs.~(1b) and~(1c) are well established~\cite{pdg}. Basing on
these observations, we believe that if one uses the existing data
to motivate the spectrum~\eqref{1} for the light vector mesons
then, staying within the same or even better accuracy, one should
accept the Regge-like interpolating formula~\eqref{1} for the
heavy vector mesons as well.

Our aim is to find holographically the functions $A(m)$ and $B(m)$
($m$ is the quark mass) in the assumption of the linear
spectrum~\eqref{1}.

The paper is organized as follows. In Sect.~2, we give a sketchy
view of the soft-wall (SW) model~\cite{son2} --- a bottom-up
holographic model which is accommodated for the description of the
Regee-like spectrum~\eqref{1}. We briefly discuss also some
attempts to apply this model to the heavy quarkonia. In Sect.~3,
we reformulate the SW model in such a way that the dependence on
the quark mass can be easily incorporated and derive our result.
The phenomenological tests are discussed in Sect.~4. A particular
example of the proposed class of models is given in Sect.~5. We
conclude in Sect.~6.

\section{Soft-wall model}

The simplest SW holographic model of Ref.~\cite{son2} describing the
unflavored vector mesons is given by the action
\begin{equation}
\label{2}
S=\int d^4\!x\,dz\sqrt{g}\,e^{-az^2}\left(
-\frac{1}{4g_5^2}F_{MN}F^{MN}\right),
\end{equation}
where
\begin{equation}
\label{3}
g=|\text{det}g_{MN}|, \qquad F_{MN}=\partial_M V_N-\partial_N V_M,
\end{equation}
and $M,N=0,1,2,3,4$. The action~\eqref{2} is defined in the AdS$_5$ background space,
the commonly used parametrization of its metrics reads
\begin{equation}
\label{4}
g_{MN}dx^Mdx^N=\frac{R^2}{z^2}(\eta_{\mu\nu}dx^{\mu}dx^{\nu}-dz^2).
\end{equation}
Here $\eta_{\mu\nu}=\text{diag}(1,-1,-1,-1)$ and $R$ denotes the
AdS$_5$ radius. Below we set $R=1$ for simplicity. The holographic
coordinate $z>0$ has the physical interpretation of inverse energy
scale. The UV boundary $z=0$ in~\eqref{4} represents the 4D
Minkovski space. The 5D vector field $V_M(x,z)$ is dual to the 4D
conserved vector current $j_\mu=\bar{\psi}_q\gamma_\mu\psi_q$ for
any quark flavor $q$. Here the precise sense of duality consists
in the identification of the boundary value $V_M(x,0)$ with the
source for the operator $j_\mu$~\cite{witten,gub}. According to
the AdS/CFT prescriptions~\cite{witten,gub}, the 5D mass of the
field $V_M$ is $m_5^2=(i-j)(i+j-4)=0$, where $j=1$ denotes, in the
given case, the spin and $i=3$ means the canonical dimension of
$j_\mu$. The ensuing gauge invariance of the action~\eqref{2}
allows to choose a convenient gauge for calculating the mass
spectrum. This is the axial gauge
\begin{equation}
\label{5}
V_z=0.
\end{equation}
The mass spectrum can be found either by calculating the vector correlator
following the AdS/CFT dictionary~\cite{witten,gub} or by finding the normalizable
solutions of the equation of motion. For our purposes, the latter method is more
convenient. The corresponding equation is
\begin{equation}
\label{6}
\partial_z\left(\frac{e^{-az^2}}{z}\partial_z v_n\right)+M_n^2\frac{e^{-az^2}}{z}v_n=0,
\end{equation}
where the eigenfunctions $v_n(z)$ stem from the Kaluza-Klein decomposition
\begin{equation}
\label{7}
V_{\mu}(x,z)=\sum_{n=0}^{\infty}V_{\mu}^{(n)}(x)v_n(z).
\end{equation}
The transverse part of $V_{\mu}^{(n)}(x)$ correspond to the 4D
physical vector fields, the index $n$ is identified with the
radial number. Writing the equation of motion for the 4D Fourier
transform $V_\mu(q,z)$, one finds the mass spetrum as the
eigenvalues $q_n^2=M_n^2$. The substitution
\begin{equation}
\label{8}
v_n=\sqrt{z}e^{az^2/2}\psi_n,
\end{equation}
transforms Eq.~\eqref{6} into the Schr\"{o}dinger form
\begin{equation}
\label{9}
-\partial_z^2\psi_n+U(z)\psi_n=M_n^2\psi_n,
\end{equation}
\begin{equation}
\label{10}
U(z)=\frac{3}{4z^2}+a^2z^2,
\end{equation}
The form of the "potential"~\eqref{10} is a consequence
of the choice of the 5D exponential background in the action~\eqref{2}.
This choice leads to a particularly simple Regge-like spectrum
\begin{equation}
\label{11}
M_n^2=4|a|(n+1).
\end{equation}
In spite of its simplicity, the predicted equality of the slope
and intercept is very close to our fit for the linear
$\omega$-meson trajectory in Table~2.

The spectrum~\eqref{11} does not depend on the 5D coupling $g_5$.
But this coupling enters the expression for the electromagnetic
decay constant\footnote{In Ref.~\cite{son1}, the constant $F$ has
the dimension of mass squared. We prefer another normalization ---
$F$ in~\eqref{12} has the dimension of mass.}~\cite{son1}
\begin{equation}
\label{12}
F_n^2=\frac{2\sqrt{2}|a|}{g_5}.
\end{equation}
After calculation of the leading contribution to the two-point
correlator of the vector currents the coupling $g_5$ can be fixed
from matching to the corresponding QCD result~\cite{son1,pom},
\begin{equation}
\label{13}
g_5^2=\frac{12\pi^2}{N_c}.
\end{equation}

We mention some attempts to accommodate the SW model for the
description of the charmonium. They can be classified as the
"shifted" models~\cite{grig} and the "rescaled"
modes~\cite{kim,fujita}. A bit simplifying the matter, in the
"shifted" models, one adds a constant $c^2$ to the
"potential"~\eqref{10} which leads to the shift of the
spectrum~\eqref{11},
\begin{equation}
\label{14}
M_n^2=4|a|(n+1)+c^2.
\end{equation}
In the "rescaled" models, one just rescales the slope parameter
$a$ in~\eqref{11}: $a\rightarrow a'$. The mass of the ground
$\psi$-meson can be reproduced by choosing the parameter $c$ or
$a'$. But as is clearly seen from Table~2, such simplistic models
fail to describe correctly the radial spectrum since the slope and
the intercept must grow simultaneously with increasing the quark
mass. In those papers, however, this circumstance was not
considered as a drawback because the aim of the proposed models
was to describe holographically the finite-temperature effects on
the $J/\psi$ meson. A much more complicated model of
Ref.~\cite{schmidt} is aimed at the description of the whole meson
spectroscopy. This model can be referred to as "shifted" one since
it results in an analytic expression for the shift $c^2$
in~\eqref{14} as a function of the quark masses and the binding
energy, with the slope being fixed. We will construct a quite
different model, in which both the slope and the binding energy
represent growing functions of the quark mass.

\section{No-wall approach and quark masses}

The $z$-dependent exponential background of the SW model~\eqref{2}
was inserted by hands with the aim of providing the Regge-like
mass spectrum~\eqref{11}. There is an alternative way to achieve
this goal --- the so-called no-wall
approach~\cite{Regge:hol,nowall}. Here one starts from the pure
AdS$_5$ space and tries to restore the UV contributions to the 5D
Lagrangian using some QCD inputs in the form of various QCD
operators. Then one guesses a IR continuation of the introduced
contributions. This step replaces guessing the 5D background. At
the end, the full contribution in the $0<z<\infty$ range can be,
in principle, effectively rewritten as a $z$-dependent background
by a certain transformation of the 5D fields~\cite{nowall}.

As the starting point for our analysis we consider the no-wall
approach~\cite{Regge:hol}. This approach is simpler for our
purposes than the original SW one because the equations of motion
for the scalar fields considered below have simple polynomial
solutions in the pure AdS$_5$ space. In the case of the SW model,
the solutions would be expressed via some trigonometric functions,
although the final qualitative conclusions would be the same. The
holographic action of the no-wall model is defined by
\begin{equation}
\label{15}
S=\int d^4\!x\,dz\sqrt{g}\left\{\sum_i\left(|DX_i|^2-m_i^2|X_i|^2\right)
-\frac{1}{4g_5^2}F_{MN}^2\right\},
\end{equation}
where $F_{MN}$ is given in~\eqref{3} and the covariant derivative
is
\begin{equation}
\label{16}
D_M X_i=\partial_M X_i - iV_M X_i.
\end{equation}
On the UV boundary, $z=0$, the scalar fields $X_i$ are identified
with sources of various QCD operators with canonical dimension
$i$. The corresponding 5D masses are~\cite{witten,gub}
\begin{equation}
\label{17}
m_i^2=i(i-4).
\end{equation}
By assumption, the fields $X_i$ acquire the $z$-dependent vacuum
expectation values $\langle X_i\rangle$ which represent the
$x$-independent solutions of the equation of motion,
\begin{equation}
\label{18}
\partial_z\left(\frac{1}{z^3}\partial_z X_i\right)=
\frac{m_i^2}{z^5}X_i,
\end{equation}
with the UV boundary condition
\begin{equation}
\label{19}
\left.\langle X_i\rangle\right|_{z=0}=0.
\end{equation}

The analogue of Eq.~\eqref{6} for the vector physical modes is
\begin{equation}
\label{20}
\partial_z\left(\frac{1}{z}\partial_zv_n\right)+\frac{M_n^2}{z}v_n=
\frac{2g_5^2}{z^3}v_n\sum_i\langle X_i\rangle^2.
\end{equation}
The change of variables $v_n=\sqrt{z}\,\psi_n$ brings this equation
into the Schr\"{o}dinger form
\begin{equation}
\label{21}
-\partial_z^2\psi_n+\left(\frac{3}{4z^2}+2g_5^2f(z)\right)\psi_n = M_n^2\psi_n,
\end{equation}
where
\begin{equation}
\label{22}
f(z)=\frac{1}{z^2}\sum_i\langle X_i\rangle^2
\end{equation}
determines the holographic "potential".

Let us consider the dimension-two operator, $i=2$, and neglect all
others. The solution of Eq.~\eqref{18} satisfying~\eqref{19} reads
\begin{equation}
\label{23}
\langle X_2\rangle=C_2^{(1)}z^2+C_2^{(2)}z^2\ln z.
\end{equation}
If we set $C_2^{(2)}=0$, the equation~\eqref{21} coincide
with~\eqref{9}, i.e. such a no-wall model looks equivalent to the
SW model. This equivalence can be explicitly shown by a
redefinition of the vector field~\cite{nowall} (see
also~\cite{schmidt}).

It should be emphasized once more that the potential in
Eq.~\eqref{21} is written near the AdS boundary, $z=0$, where, by
assumption, the holographic correspondence allows to use the QCD
inputs. In order to obtain the mass spectrum we need to continue
the function $f(z)$ to the infrared domain, $z\rightarrow\infty$.
The linear spectrum of the kind~\eqref{11} can be obtained only if
\begin{equation}
\label{24}
f(z)|_{z\rightarrow\infty}=a^2z^2.
\end{equation}
The fact that the UV asymptotics of the 5D field dual to the dim2
operator provides automatically the correct IR asymptotics is a
lucky coincidence taking place for the simplest SW model. In the
general case, these asymptotics are different even for the field
$X_2$, see Appendix. In principle, the sum in~\eqref{22} may lead
to the asymptotics~\eqref{24} even in the absence of the dim2
operator. But the matter looks as if the dim2 operator were dual
to the sum~\eqref{22} in the infrared: Whether we introduce the
dim2 operator and neglect all others or we deal with the whole
sum~\eqref{22} assuming the asymptotics~\eqref{24}. We remind the
reader that the dim2 operator can be built in QCD --- this is the
gauge non-invariant gluonic operator $A_\mu A^\mu$. Its vacuum
expectation value (v.e.v.) $\langle A_\mu A^\mu\rangle$ often
serves (in the Landau gauge where it is minimal) for a
parametrization of some important non-perturbative
effects~\cite{dim2}. There are arguments~\cite{zakharov} that
$\langle A_\mu A^\mu\rangle$ should emerge from a resummation of
perturbative corrections to the unit operator in the Operator
Product Expansion (OPE) of the correlation functions~\cite{svz},
i.e. one either deals with the infinite sum of these corrections
or with $\langle A_\mu A^\mu\rangle$. In some sense, we have a
holographic analogue for such kind of duality.

Our aim is to extract the dependence of the linear mass spectrum
on the quark mass. Such a dependence can appear only from the
v.e.v. $\langle X_3\rangle$ since the field $X_3$ is dual to the
quark bilinear operator $\bar{q}q$. The quark mass emerges from
the AdS/CFT prescription derived in Ref.~\cite{kleb} which states
that the solution of classical equation of motion for a scalar
field $\Phi$ corresponding to an operator $O$ of canonical
dimension $i$ has the following form near the 4D boundary
$z\rightarrow0$,
\begin{equation}
\label{25}
\Phi(x,z)\rightarrow
z^{4-i}\left[\Phi_0(x)+\mathcal{O}(z^2)\right]+z^{i}\left[\frac{\langle
O(x)\rangle}{2i-4}+\mathcal{O}(z^2)\right],
\end{equation}
where $\Phi_0(x)$ acts as a source for $O(x)$ and $\langle
O(x)\rangle$ denotes the corresponding condensate. In QCD, the
quark mass $m$ acts as the source for the operator $\bar{q}q$.

For the canonical dimension $i=3$, the solution of Eq.~\eqref{18}
satisfying~\eqref{19} is $\langle
X_3\rangle=C_3^{(1)}z+C_3^{(2)}z^3$. According to the
prescription~\eqref{25}, this solution can be rewritten in terms
of the physical quantities. In our isospin-zero case, the
corresponding expression is
\begin{equation}
\label{26}
\langle X_3\rangle=\xi m z+\frac{\sigma}{2\xi}z^3,
\end{equation}
where $\sigma$ denotes the quark condensate $\langle
\bar{q}q\rangle$ and the normalization factor $\xi$ was calculated
in Ref.~\cite{cohen},
\begin{equation}
\label{27}
\xi^2=\frac{N_c}{4\pi^2}.
\end{equation}

If we take into account the field $X_3$ and neglect all other
scalar fields, the action of the model will coincide with the
action of the original bottom-up models~\cite{son1,pom} in the
vector sector. The field $X_3$ is usually exploited for the
holographic description of the chiral symmetry breaking in QCD.
This phenomenon is important in the axial-vector sector which we
do not consider. For our purposes, $X_3$ is crucial to derive the
dependence of the vector spectrum on the current quark masses.

The relation~\eqref{26} allows to extract explicitly the
$m$-dependent terms in~\eqref{22},
\begin{equation}
\label{28}
f(z)\rightarrow \xi^2m^2+m\sigma z^2+\tilde{f}(z),
\end{equation}
where the contribution $\frac{\sigma^2}{4\xi^2}z^6$ is absorbed
into the new sum $\tilde{f}(z)$. In order to reproduce the Regge
form of the spectrum the first two terms of the IR asymptotics of
$\tilde{f}(z)$ must be given by
\begin{equation}
\label{29}
\left.\tilde{f}(z)\right|_{z\rightarrow\infty}=a^2z^2+\delta.
\end{equation}

If we knew explicitly all coefficient in the UV expansion of
$\tilde{f}(z)$ the summation would lead to a definite function
with the IR asymptotics~\eqref{29}. An illustrative example is
given in Appendix. In the real situation, we need to exploit some
interpolation scheme. We make use of the following simplification:
The function $\tilde{f}(z)$ is replaced by its IR
asymptotics~\eqref{29}. The UV asymptotics of the holographic
potential in~\eqref{21} is controlled by the first term
$\frac{3}{4z^2}$. The UV asymptotics and the behavior at
intermediate $z$ of $\tilde{f}(z)$ is interpolated by the constant
$\delta$ in~\eqref{29}. The example of a solvable model in
Appendix demonstrates that such a simplification should not modify
strongly the final spectrum. This partly justifies our
simplification.

Thus, we arrive at the following equation on the mass spectrum,
\begin{equation}
\label{30}
-\partial_z^2\psi_n+\left[\frac{3}{4z^2}+2g_5^2(\sigma m+a^2)z^2+
2g_5^2\xi^2m^2+2g_5^2\delta\right]\psi_n = M_n^2\psi_n.
\end{equation}
The factor at $z^2$ in~\eqref{30} determines the slope of the
linear trajectory. It has two contributions and one of them
depends on the quark mass. This situation correlates with the QCD
sum rules~\cite{svz,reinders} in the large-$N_c$ limit~\cite{sr},
in which the slope is determined by the v.e.v.'s of the dim4
operators. The OPE contains two such v.e.v.'s~\cite{svz}:
$m\langle\bar{q}q\rangle$ and $\frac{\alpha_s}{\pi}G_{\mu\nu}^2$.
In the light quark sector, the gluon condensate dominates while in
the heavy quark one, the first v.e.v. is dominant.

The spectrum given by Eq.~\eqref{30} is
\begin{equation}
\label{31}
M_n^2=4\sqrt2\,g_5\sqrt{\sigma m+a^2}(n+1)+2g_5^2\xi^2m^2+2g_5^2\delta.
\end{equation}
This expression shows the parametric dependence of the linear
spectrum on the quark mass $m$. After obvious redefinition of the
constants, the spectrum~\eqref{31} may be written in a more
compact form,
\begin{equation}
\label{32}
M_n^2=\sqrt{\alpha+\beta m}(n+1)+\gamma m^2+\delta.
\end{equation}
The value of the constant $\gamma$ follows directly from the
relations~\eqref{13},~\eqref{27} and~\eqref{31},
\begin{equation}
\label{33}
\gamma=2g_5^2\xi^2=6.
\end{equation}
In the next Section, we will use this value for some
phenomenological estimates.

\section{Phenomenological tests}

In the limit $m\rightarrow0$, the parameters $\alpha$ and $\delta$
in the spectrum~\eqref{32} can be fixed from the $\omega$-meson
trajectory (Table~2), $\sqrt{\alpha}\approx1$~GeV$^2$,
$\delta\approx0$. The remaining parameter $\beta$ and the quark
masses may be estimated from the fits in Table~2. The charmonium
trajectory gives $\beta\approx7$~GeV$^3$, $m_c\approx1.1$~GeV. The
bottomonium one leads to $\beta\approx8$~GeV$^3$,
$m_b\approx3.9$~GeV. At the scale 2~GeV, the Particle Data cites
the current quark masses in the $\overline{\text{MS}}$ scheme
$m_c=1.27$~GeV, $m_b=4.18$~GeV~\cite{pdg}. It should be noted that
the $b$-quark mass is expected to be lower in the real excited
bottomonia since the energy scale is about 5~GeV per quark. The
renormalization group running of the quark masses~\cite{pdg}
predict that (at the one-loop level) the cited value for $m_b$
must be then divided by a factor of $1.2$. In any case, taking
into account the rough approximations which we have used, the
overall agreement with the experimental data is not bad. This
means that the predicted value of the parameter
$\gamma$~\eqref{33} --- the rate of the squared meson mass
dependence on $m^2$ --- lies close to the phenomenologically
acceptable range. A global fit of the data gives a value of
$\gamma$ in the range $5\lesssim \gamma \lesssim 7$ depending on
the assumptions and inputs.

Considering the limit $m\rightarrow0$ analytically, the mass
relation~\eqref{31} results in a certain linear in $m$ correction
to the slope,
\begin{equation}
\label{34}
4\sqrt2\,g_5\sqrt{\sigma m+a^2}=4\sqrt2\,g_5a+\frac{2\sqrt2\,g_5}{a}\sigma m+\mathcal{O}(m^2).
\end{equation}
Assuming the linear form of the spectrum at $m=0$, the consistency
of QCD sum rules in the large-$N_c$ limit (the so-called planar
sum rules) leads to the slope~\cite{sr}
\begin{equation}
\label{35}
4\sqrt2\,g_5a=\frac{48\pi^2}{N_c}f_\pi^2=2m_\rho^2,
\end{equation}
where $f_\pi$ is the weak pion decay constant,
$f_\pi=92.4$~MeV~\cite{pdg}, and $m_\rho$ denotes the $\rho$ or
$\omega$ meson mass. The expansion~\eqref{34} combined
with\footnote{As an alternative input we could use the so-called
KSFR relation $F_\rho^2=2f_\pi^2$ which also follows from the
planar QCD sum rules~\cite{sr}. Matching this relation with the
expression~\eqref{12} leads to the same result.}~\eqref{35} yields
the slope
\begin{equation}
\label{36}
4\sqrt2\,g_5\sqrt{\sigma m+a^2}=\frac{48\pi^2}{N_c}f_\pi^2+
\frac{4\sigma m}{f_\pi^2}+\mathcal{O}(m^2)=2m_\rho^2-2m_\pi^2+\mathcal{O}(m^2),
\end{equation}
where we have used the Gell-Mann--Oakes--Renner (GOR) relation
$m_\pi^2f_\pi^2=-2m\sigma$. The dual Veneziano-like
amplitudes~\cite{ampl} predict the same correction~\eqref{36} to
the slope when the pion mass is taken into account. In this sense,
our holographic model passes one more test.

It should be remarked that all relations obtained within the
planar QCD sum rules can be derived in the bottom-up holographic
models since, in some sense, the bottom-up approach represents
just a compact 5D language for expressing the phenomenology of
these sum rules~\cite{rewr}. We have taken~\eqref{35} as an
external input in order not to complicate the matter. The same can
be said about the GOR relation which can be reproduced in the
bottom-up models if the axial-vector field is
introduced~\cite{son1,pom}.

The possible additional tests may follow from the calculation of
the vector correlator. Its high-momentum euclidean asymptotics can
be matched with the OPE in QCD~\cite{svz}. In our case, some
specific polynomial and logarithmic contributions will appear due
to the quark masses. Similar contributions emerge in OPE if the
non-zero quark masses in the fermion loops are taken into account.
Unfortunately, we could not find the corresponding trustful
results in the literature. In particular, the calculations of the
quark mass contributions to the unit operator (the leading
logarithm) presented in~\cite{reinders} and~\cite{jamin} do not
match.

The non-relativistic potential models give their relations for the
meson masses in the form $M=m_1+m_2+E$, where $m_1$, $m_2$ are the
quark masses and $E$ is the binding energy. Our result~\eqref{32}
shows that the binding energy grows linearly with the quark mass
in the heavy quark limit, $m\rightarrow\infty$, due to~\eqref{33}
(where $\gamma>4$) . This behavior is in a qualitative agreement
with the experimental data~\cite{pdg}. For example, in the case of
the ground states one has $M_{\psi}-2m_c\approx0.56$~GeV and
$M_{\Upsilon}-2m_b\approx1.1$~GeV if $m_{c,b}$ are taken at 2~GeV
(taking $m_b$ at 5~GeV leads to $M_{\Upsilon}-2m_b\approx2.5$~GeV)
. The given effect can be easily interpreted: When a
non-relativistic quark of mass $m$ is created and moves with the
velocity $v$, the binding energy should compensate its kinetic
energy $\frac{mv^2}{2}$.

\section{A gauge non-invariant example}

The mass relation~\eqref{31} (or~\eqref{32}) is not just a result
of some particular model, rather it represents a result given by a
class of holographic models. Within this class, the difference in
the final expression for the mass spectrum competes with (or
exceeds) the accuracy of the method, so that the predictions can
be regarded as equivalent. This class can be extended if we do not
impose the requirement of the 5D gauge invariance and consider the
relation~\eqref{5} as a part of the definition for the 4D physical
modes. Then it is easy to construct such a fine-tuning that the
sum $f(z)$~\eqref{22} contains a finite number of terms. This is
equivalent to considering only a few of operators of the lowest
dimensions in the OPE~\cite{svz} for the calculations of the
hadron masses (the standard approximation in the QCD sum
rules~\cite{svz,reinders}). In the given Section, we demonstrate a
typical example for this kind of models.

The action of the model is
\begin{equation}
\label{37}
S=\int
d^4\!x\,dz\sqrt{g}\left\{\mathcal{L}_\text{V}+\mathcal{L}_\text{S}+\mathcal{L}_\text{int}\right\},
\end{equation}
where
\begin{equation}
\label{38}
\mathcal{L}_\text{V}=-\frac{1}{4g_5^2}F_{MN}F^{MN},\qquad
F_{MN}=\partial_MV_N-\partial_NV_M;
\end{equation}
\begin{equation}
\label{39}
\mathcal{L}_\text{S}=\frac12\sum_{k=1}^{3}\left(\partial_M\varphi_k\partial^M\varphi_k-m_i^2\varphi_k^2\right);
\end{equation}
\begin{equation}
\label{40}
\mathcal{L}_\text{int}=\frac12V_MV^M\left(g_1\varphi_1+g_2\varphi_2^2+g_3\varphi_3\right).
\end{equation}

We choose the following correspondence between the 5D fields and
the operators in QCD,
\begin{equation}
\label{41}
\varphi_1\longleftrightarrow G_{\mu\nu}^2,\qquad
\varphi_2\longleftrightarrow \bar{q}q,\qquad
\varphi_3\longleftrightarrow (\bar{q}q)^2.
\end{equation}
The canonical dimensions of the operators in~\eqref{41} are
\begin{equation}
\label{42}
i_1=4,\qquad i_2=3,\qquad i_3=6,
\end{equation}
which according to~\eqref{17} dictate the masses
\begin{equation}
\label{43}
m_1^2=0,\qquad m_2^2=-3,\qquad m_3^2=12.
\end{equation}
The solutions of Eq.~\eqref{18} satisfying the boundary condition~\eqref{19} read
\begin{equation}
\label{44}
\varphi_1(z)=C_1z^4,\qquad
\varphi_2(z)=C_{21}z+C_{22}z^3,\qquad
\varphi_3(z)=C_3z^6.
\end{equation}
Writing $\varphi_2$ in the form~\eqref{26} we arrive at the
following $z$-dependent vector mass term
\begin{equation}
\label{45}
m_V^2(z)=g_5^2\left[g_2\xi^2m^2z^2+(g_1C_1+g_2\sigma
m)z^4+\left(g_2^2\frac{\sigma^2}{4\xi^2}+g_3C_3\right)z^6\right].
\end{equation}
The analogue of Eq.~\eqref{21} becomes
\begin{equation}
\label{46}
-\partial^2_z\psi_n+\left[\frac{3}{4z^2}+\frac{m_V^2}{z^2}\right]\psi_n=M_n^2\psi_n.
\end{equation}
In order to reproduce the linear spectrum we must set to zero the
coefficient at $z^6$ in $m_V^2(z)$~\eqref{45}. Then the
"potential" of Eq.~\eqref{46} can be reparametrized as
\begin{equation}
\label{47}
U(z)=\frac{3}{4z^2}+\frac14(\alpha+\beta m)z^2+\gamma m^2.
\end{equation}
This parametrization leads to the mass spectrum~\eqref{32} with $\delta=0$.

\section{Conclusions}

The experimental data on the excited heavy and light unflavored
vector quarkonia strongly suggests that the spectrum follows the
Regge behavior $M_n^2=An+B$, here $n$ means the radial quantum
number. Both the slope $A$ and the intercept $B$ grow rapidly with
increasing the quark mass. In the existing bottom-up holographic
models, this behavior is not reproduced since either $A$ or $B$ is
independent of the quark mass. We have constructed a bottom-up
holographic model in which both $A$ and $B$ represent growing
functions of the quark mass. Our model passes qualitatively
several phenomenological tests.\\
i) In the heavy-quark limit, the behavior is $A\sim\sqrt{m}$,
$B=\gamma m^2+\mathcal{O}(\sqrt{m})$, where $m$ is the mass of
quarks constituting the unflavored vector meson. This explains
qualitatively (and even semiquantitatively) why the
intercept grows with $m$ much faster than the slope. \\
ii) The considered holographic models predict $\gamma=6$ in the
asymptotics for $B$. The given value agrees well with the
phenomenology. In addition, it predicts growing the quarkonium
binding energy with the quark mass. This effect is clearly seen in
the experimental data. \\
iii) In the limit $m\rightarrow0$, the correction to the linear
spectrum due to $m\neq0$ is consistent with the Veneziano-like
dual amplitudes.

The presented approach can be applied to various physical
problems. For instance, it is possible to analyze the impact of
the finite-temperature effects on the whole excited spectrum of
the heavy vector mesons.

\section*{Acknowledgments}

The work was partially supported by the Saint Petersburg State
University grants ¹ 11.38.660.2013 and 11.48.1447.2012, by the
RFBR grant 13-02-00127-a and by the Dynasty Foundation.

\section*{Appendix}

In this Appendix, we give an illustrative example for some
statements made in Sect.~3. The first statement was that in the 5D
holographic action, the UV part of asymptotics restored by the
method of Sect.~3 generically does not coincide with the
corresponding IR asymptotics even for the scalar field dual to the
dim2 operator.

Consider the SW model~\eqref{2}. It leads to the linear
spectrum~\eqref{11}. If we wish to have an arbitrary intercept,
$$
M_n^2=4|a|(n+1+b)
\eqno (A.1)
$$
where $b$ is a free intercept parameter (this form is more
convenient than~\eqref{14}), the dilaton background of the
action~\eqref{2} must be modified in the following way~\cite{genSW},
$$
S=\int d^4\!x\,dz\sqrt{g}\,e^{-az^2}U^2(b,0;az^2)\left(
-\frac{1}{4g_5^2}F_{MN}F^{MN}\right).
\eqno (A.2)
$$
Here $U$ denotes the Tricomi confluent hypergeometric function.
The change of the vector field
$$
V_M=e^{az^2/2}U^{-1}(b,0;az^2)\tilde{V}_M
\eqno (A.3)
$$
transforms the dilaton background into an effective $z$-dependent
mass term,
$$
S=\int
d^4\!x\,dz\sqrt{g}\left\{-\frac{1}{4g_5^2}\tilde{F}_{MN}\tilde{F}^{MN}+\right.
$$
$$
\left.
\frac{a^2}{2g_5^2}\left[z^2+2\frac{b}{a}-\frac{2U(b-1,0;az^2)}{aU(b,0;az^2)}\right]^2\tilde{V}_M\tilde{V}^M\right\}.
\eqno (A.4)
$$
The UV asymptotics of the mass term results from the Taylor
expansion
$$
\left.z^2+2\frac{b}{a}-\frac{2U(b-1,0;az^2)}{aU(b,0;az^2)}\right|_{z\rightarrow0}=
z^2\left[1+2b\left(\ln(az^2)+c\right)\right],
\eqno (A.5)
$$
$$
c=2\gamma-1+\psi(b)+b\left[\psi(b+1)-\psi(b)\right],
$$
where $\psi$ denotes the digamma function and $\gamma$ is the
Euler constant. The expression $(A.5)$ shows that the UV
asymptotics can be reproduced by the method of Sect.~3 if for the
contribution $\langle X_2\rangle$ in~\eqref{22} the whole
solution~\eqref{23} is used. Setting $C_2^{(2)}=0$ is equivalent
to setting $b=0$. Exactly this case was considered in
Ref.~\cite{nowall} where the no-wall holographic model was
proposed.

The IR asymptotics follows from
$$
\left.z^2+2\frac{b}{a}-\frac{2U(b-1,0;az^2)}{aU(b,0;az^2)}\right|_{z\rightarrow\infty}=
-\left(z^2+2\frac{b}{a}\right)+\mathcal{O}(z^{-2}).
\eqno (A.6)
$$
We see that if $b\neq0$ the UV and IR asymptotics of the effective
mass term are different. This proves our statement.

If the IR asymptotics $(A.6)$ is substituted to $(A.4)$ and only
the two leading terms are retained, the spectrum of the ensuing
model will coincide with $(A.1)$. This illustrates the reason why
in~\eqref{29} only the two leading contributions have been
considered.

The expansion of the r.h.s. of $(A.5)$ models the UV contributions
restored by the method of Sect.~3. It is clear that if all
coefficients in the UV part are known, the UV contributions can be
summed into a certain function --- here the l.h.s. of $(A.5)$ ---
and then continued to the IR region, $z\rightarrow\infty$. But if
we know the form of the spectrum to be obtained, it is enough to
guess the leading and next-to-leading IR contributions. This
property we have exploited in Sect.~3.

\end{document}